# The impact of social status on the formation of collaborative ties and effort provision: An experimental study[1]

Gergely Horvath[2] and Mofei Jia[3]

## Abstract

We study whether competition for social status induces higher effort provision and efficiency when individuals collaborate with their network neighbors. We consider a laboratory experiment in which individuals choose a costly collaborative effort and their network neighbors. They benefit from their neighbors' effort and effort choices of direct neighbors are strategic complements. We introduce two types of social status in a 2x2 factorial design: 1) individuals receive monetary benefits for incoming links representing popularity; 2) they receive feedback on their relative payoff ranking within the group. We find that link benefits induce higher effort provision and strengthen the collaborative ties relative to the Baseline treatment without social status. In contrast, the ranking information induces lower effort as individuals start competing for higher ranking. Overall, we find that social status has no significant impact on the number of links in the network and the efficiency of collaboration in the group.

**JEL codes:** D85, D62, C92
**Keywords:** network formation, collaboration, social status, efficiency, experiment

---

[1] Gergely Horvath acknowledges support for this research from the National Natural Science Foundation of China (NSFC), under the 2021 NSFC Research Fund for International Excellent Young Scientists (RFIS-II) program, grant nr. 72150610501. Mofei Jia acknowledges support for this research from the National Natural Science Foundation of China (NSFC), under the NSFC Young Scientists Fund, grant nr. 72203174, and also acknowledges support from Xi'an Jiaotong-Liverpool University under the Research Enhancement Fund REF-20-01-03.

[2] Division of Social Sciences, Duke Kunshan University. Address: Division of Social Sciences, Duke Kunshan University, No. 8 Duke Avenue, Kunshan, Jiangsu Province, China 215316. E-mail address: horvathgergely@gmail.com. Tel: +86 155-01683821.

[3] *Corresponding author.* International Business School Suzhou, Xi'an Jiaotong-Liverpool University. Address: International Business School Suzhou, Xi'an Jiaotong Liverpool University, No. 111 Ren'ai Rd., Suzhou Industrial Park, 215123 Suzhou, China. E-mail address: mofei.jia@gmail.com. Tel: +86 159-62448710.

# 1. Introduction

Collaboration often occurs in a social network, whereby individuals or companies work on projects with their direct network neighbors who also collaborate with their neighbors, and so on. Typical examples include co-authorship networks (Cassi et al., 2012; Ductor et al., 2014; Goyal et al. 2006; Juhász et al., 2020) and R&D collaboration among firms and academic institutions (Dawid and Hellmann, 2014; Goyal and Moraga-Gonzalez, 2001; König et al., 2012; Leydesdorff and Wagner, 2008; Rank et al., 2006, Zu et al., 2011). Collaboration in these settings has the features that agents benefit from the effort of their direct neighbors and also have incentives to exert higher effort when their neighbors do so. In other words, positive externalities and strategic complementarity in effort choices exist between network neighbors. It is well-known that positive externalities make the outcomes of individual utility-maximizing choices socially inefficient. In our case, where individuals choose effort and their network neighbors, effort provision and network connectivity may be lower in the decentralized equilibrium than in the social optimum (Hiller, 2017). Experimental evidence using the linear-quadratic utility function to capture the incentives show that albeit individuals overprovide effort relative to the equilibrium level in small (but not in large) groups (Gallo and Yan, 2021), they form too few links which reduces efficiency to a great extent (Horvath, 2023).

In this paper, we ask whether incentives derived from status seeking can increase efficiency in a setting when individuals choose collaborative efforts and their network neighbors. The motivation comes from the observation that collaborators working on innovative projects typically gain social status from completing those projects, besides the direct monetary benefits. For example, academics gain social status from publishing in top journals, and companies gain social recognition from discovering new technologies and marketing new products. Status may lead to monetary benefits, such as attracting grants and investors, or to non-monetary benefits, such as recognition in the community. These additional benefits may provide incentives for individuals to exert higher effort and form more collaborative ties. We thus ask how status seeking affects the extent of effort provision, the formation of collaborative ties, and the welfare that collaboration brings.

We study these questions in a controlled laboratory setting, which allows us to manipulate different types of social status. In the experiment, participants choose their network neighbors and a costly effort level, that represents the activity individuals exert on collaborative projects. They make these decisions for 40 periods in a fixed group of 5 individuals. The network links are formed in a unilateral way. The monetary gains of collaboration are represented by the linear-quadratic utility function that captures the positive externality of individual effort on neighbors in the network, as well as, the complementarity of effort between collaborating individuals. In the Baseline treatment, this is the only benefit individuals gain from their choices. We introduce status benefits to this setting in two different forms, using a 2x2 factorial design. Firstly, individuals obtain additional monetary status benefits for incoming links, which represent gains derived from popularity. We label this treatment as Link benefit. Van Leeuwen et al. (2020) implemented a similar setting in a network formation game of local public goods provision. They show that status benefits induce individuals to exert higher effort (that is, to provide more local public goods) in order to attract more links, which leads to the efficient provision of public goods and the formation of a star network, i.e., the efficient configuration in their setting. We expect a similar effect, whereby individuals exert more collaborative efforts that leads to more links being formed due to the existing complementarities, and eventually to the formation of a complete network, which is the efficient network structure in our setting. Secondly, we induce status competition by providing information on the relative ranking of the individuals within the group in terms of payoffs earned. Extensive literature shows that individuals care about their relative position and try to achieve a higher position relative to their reference group (Fisher and Torgler, 2013; Frank, 1985, 2005; Senik, 2004; Weiss and Freshman, 1998). They may do so in our game by increasing their effort levels and forming more beneficial links that rises their payoffs. We label this treatment as Ranking. In a fourth Interaction treatment, we include both types of social status.

We find in our Baseline treatment that individuals do not form a Nash equilibrium network and create less links than it is prescribed by the efficient complete network. They overprovide effort relative to the Nash equilibrium level on the network they formed. The efficiency level in the group falls short of the total payoffs that could be obtained in the complete network with equilibrium and efficient effort levels. These findings replicate the previous results found in Horvath (2023) and validate our experiment. Turning to the impact of social status, we find that individuals provide significantly higher effort in the Link benefit than in the Baseline treatment. However, this does not lead to more links being formed. Instead, it strengthens existing links since the fraction of links initiated by both individuals involved in the link increases relative to the Baseline. Overall, the efficiency level of the Link benefit treatment is not significantly different from that of the Baseline treatment because the connectivity of the network does not change.

Turning to the impact of payoff ranking information, we find that individuals provide significantly less effort in the Ranking than in the Baseline treatment, while the number of links is not significantly different between the two treatments. Effort provision starts high, above the equilibrium level, in the Ranking treatment. However, over time, individuals ranked low in the group cut their effort provision, which allows them to achieve a higher relative ranking in the group since their effort moves closer to the individually optimal value providing higher payoffs. In contrast, individuals ranked high in the group do not change their effort choices, which leads to lower payoffs due to the negative externalities from others reducing their effort in the group. In turn, the payoff order changes and individuals ranked low keep lowering their effort levels. This dynamics leads to lower average effort provision in the last 10 periods of the game relative to the Baseline treatment. However, the difference in effort is not large enough to create a significant difference in payoffs as the number of links is not significantly different. We show that in the Interaction treatment the impact of ranking information dominates the effects of benefits received for incoming links, implying that individuals provide significantly lower effort in this treatment than in the Baseline.

In sum, we find that social status manipulations affect the effort provision, but have no impact on the number of total links formed and the efficiency in the group. The different types of social status have opposite effects on effort provision, and the negative effects of payoff ranking dominate the positive effects of benefits received for popularity. Overall, we do not find that status benefits and competition for social status would increase efficiency in our game of collaboration and network formation. This is in contrast with the results obtained for local public good provision in van Leeuwen et al. (2020). Other mechanisms are thus needed to promote collaborative efforts, such as subsidizing effort and link formation.

Our paper contributes to the experimental literature on social network games and network formation. Choi et al. (2016) provides a detailed review of this recent literature. One strand of the literature studies strategic interactions on fixed networks, including prisoner's dilemma games (Rand et al., 2014), coordination games (Berninghaus et al., 2002; Cassar, 2007), local public goods games (Caria and Fafchamps, 2020; Fatas et al., 2010; Rosenkrantz and Weitzel, 2012; Suri and Watts, 2011), games of binary actions with strategic complements and substitutes (Charness et al., 2014). Two most related papers are Gallo and Yan (2021) and Antinyan et al. (2020). Gallo and Yan (2021) study effort provision in the linear-quadratic game with positive externality and complementarity on fixed networks. They show that individuals overprovide effort relative to the Nash equilibrium level in small groups, while follow the Nash equilibrium predictions in larger groups. Our paper introduces network formation and social status effects into the same game. Antinyan et al. (2020) also study status competition in networks as we do. However, they consider a game of conspicuous consumption with negative externality and complementarity in the consumption of positional goods, following the theoretical model in Ghiglino and Goyal (2010). Their game is played on fixed networks and features the competition for social status as part of the utility function and the resulting experimental payoffs, instead of considering psychological effects that do not per se affect payoffs, as we do in our Ranking treatment.

Another part of this literature studies network formation using experiments grounded in game theory. Earlier papers (Berninghaus et al., 2007; Callander and Plot, 2005; Caria and Fafchamps, 2020; Falk and Kosfeld, 2012; Goeree et al., 2009) focus on the network formation model in Bala and Goyal (2000), in which benefits depend on the number of direct and indirect connections of the individuals, while more recent papers (Choi et al., 2022; Rong and Houser, 2015; Van Leeuween et al., 2020) study local public goods provision following Galeotti and Goyal (2010). Two most related papers are van Leeuwen et al. (2020) and Horvath (2023). Van Leeuwen et al. (2020) study status effects in local public good provision and show that monetary status benefits received after incoming links promote the efficient provision of public goods and the formation of the most efficient star network. Horvath (2023) studies the linear-quadratic network formation game featured in this paper as well. He tests the predictions of Hiller (2017) by varying the group size and linking costs across treatments and obtains that individuals overprovide effort but do not form sufficient number of links, which hurts efficiency. Our paper tests status competition as a mechanism to increase efficiency in the same game.

Finally, our paper contributes to the literature on relative concerns and status competition. A large literature shows that individuals care about their relative position compared to their reference group (see, for example, Clark and Oswald, 1996; Ferrer-i-Carbonell, 2005; Fisher and Torgler, 2013; Senik, 2004, 2008). The impact of this idea on incentivizing effort provision in a work setting has been explored by many papers (see, e.g., Eriksson et al., 2009; Azmat et al., 2019; Azmat and Irriberi, 2010; Song et al., 2018). Feedback on relative performance may encourage individuals to exert higher effort even under fixed wages (Falk and Ichino, 2006; Kuhnen and Tymula, 2012, Mas and Moretti, 2009), but it may also lead to sabotage between competitors (Charness et al., 2014) and individuals lagging behind may reduce their effort (Eriksson et al., 2009; Gill and Prowse, 2012). While most papers study the impact of relative performance feedback on individual work tasks, we introduce it in a setting where individuals choose their collaborators and their effort has positive spillovers on others. We show that in this setting, relative performance feedback leads to the reduction of effort provision.

## 2. Experimental design

### 2.1 Baseline experimental framework

Our baseline experimental treatment is based on the one-sided network formation game with a linear-quadratic utility function as introduced in Hiller (2017). For the sake of the feasibility and simplicity of the experiment, we consider a group of $N = 5$ players. Each player $i$ simultaneously decides on her effort level $x_i \in [0, 20]$ and chooses the set of other players within the group she would like to link to. Formally, player $i$ chooses $g'_{ij} \in \{0,1\}$ for each $j \in N\setminus i$, where $g'_{ij} = 1$ if player $i$ initiates a link to player $j$ and $g'_{ij} = 0$ if not. We consider unilateral link formation such that a link is established between players $i$ and $j$ in the network if either player $i$ or player $j$ (or both) initiates the link. The linking choices of all players jointly determine the network structure, that can be represented by the adjacency matrix $G$, in which $g_{ij} = 1$ if there is a link established between players $i$ and $j$, i.e., either $g'_{ij} = 1$ or $g'_{ji} = 1$ or both, and $g_{ij} = 0$ otherwise. If a link is established between players $i$ and $j$, we say that they are network neighbors.

In the experiment, we apply the following payoff function that determines the per-period earnings of the participants based on the effort and linking choices of all group members:

$$\pi_i(x, G) = 10x_i - 4x_i^2 + 0.4x_i \sum_{k \in N_i(G)} x_k - 3.9\eta_i'. \qquad (1)$$

The first term in payoff function (1) represents individual $i$'s benefit from her own effort, and the second term refers to the quadratic cost of exerting effort which results in increasing marginal costs. Positive externalities and strategic complementarity resulting from the effort choices of individual $i$ and her network neighbors are captured by the third term, where $N_i(G)$ is the set of direct neighbors of individual $i$ in the network $G$. In other words, this term implies that individual $i$ benefits from the effort of her network neighbors and has incentives to exert higher effort when her neighbors exert higher effort. The last term indicates the cost for initiating a link for individual $i$, where $\eta_i'$ stands for the number of links initiated by individual $i$, i.e., $\eta_i' = |\{j \in N: g'_{ij} = 1\}|$.

The theoretical literature on network games has extensively analyzed the introduced game with positive externalities and strategic complementarity of effort choices. Ballester et al. (2010) show that the Nash equilibrium effort level of individual $i$ is determined by the Katz-Bonacich centrality, which is the weighted sum of the individual's number of direct and indirect neighbors. Hiller (2017) shows that the equilibrium network structure belongs to the core-periphery family whereby group members can be assigned into two sets: core members who connect to everyone in the network and peripheral nodes who connect only to the core members.

Using these theoretical results and the parameter values chosen for the experiment, we show the equilibrium predictions for the effort level and network structure in the Baseline treatment in

Table 1. There are three equilibrium networks: the empty, star, and complete networks. In the empty network, no links are initiated and in the complete network all possible links are present. In the star network a single node, the center of the star, connects to everyone else, while all others connect only to the center. Among the three equilibria, individual's effort level and payoffs are highest under the complete network and smallest in the empty network. Table 1 also shows the efficient effort levels that results in the maximum total payoffs of the group under each equilibrium network structure. These effort levels are higher than the equilibrium effort levels due to the positive externalities. The total welfare is maximized in the complete network with efficient effort choices.

**Table 1: Effort level and payoffs in the Nash equilibria and the corresponding efficient allocation**

|  | Nash equilibrium | Efficient allocation |
|---|---|---|
| **Effort** | Empty: 2.5 | Empty: 2.5 |
|  | Star: (3.65, 2.86) | Star: (5.36, 3.57) |
|  | Complete: 4.17 | Complete: 12.5 |
| **Payoffs** | Empty: 12.5 | Empty: 12.5 |
|  | Star: (26.58,12.51) | Star: (26.82,13.95) |
|  | Complete: 26.92 | Complete: 54.69 |

## 2.2 Status treatments and hypotheses

As shown above, the Nash equilibrium of the game is not socially efficient due to the positive externalities. With the potential to increase efficiency, we introduce behavioral mechanisms in the form of competition for social status. In the experiment, we use a 2 x 2 between-subjects factorial design by introducing two types of status competition (see Table 2). In the first setting, individuals receive monetary status benefits for each incoming link. We label this treatment as *Link benefit*. The payoff function changes from (1) to the following:

$$\pi_i(x, G) = 10x_i - 4x_i^2 + 0.4x_i \sum_{k \in N_i(G)} x_k - 3.9\eta'_i + 6\omega'_i, \qquad (2)$$

where the last term in the payoff function (2) represents the status benefits from each incoming link, and $\omega_i'$ stands for the number of links received by individual $i$, i.e., $\omega_i' = \left|\{j \in N : g'_{ji} = 1\}\right|$. We note that the Nash equilibrium of the network formation game remains unchanged in Link benefit relative to the Baseline because the status benefits result from other individuals' linking decisions and thus does not alter an individual's incentives.

Table 2: Baseline and treatments in the experiment

| Group Size N=5 | NO linking benefit | Monetary status benefit from incoming links |
| --- | --- | --- |
| NO ranking information | Baseline | Link benefit |
| Ranking information | Ranking | Interaction |

While the Nash equilibrium of the stage game does not change, there may be other dynamic effects that alter the outcomes. The underlying idea of this status manipulation follows from the results obtained in van Leeuwen et al. (2020) who show that status benefits facilitate the provision of public goods and increase efficiency in social networks, leading to the formation of a star network, i.e., the efficient configuration in their setting. The mechanism is that status benefits provide incentives for individuals to increase effort provision in order to attract more incoming links and realize the status benefits. In other words, individuals engage in competition for incoming links by providing higher effort. In turn, in our game, the network connectedness may increase as well since individuals have incentives to link to high effort providers. As a consequence, the individual's effort provision may further increase because their best-response effort level positively depends on the sum of effort levels of their neighbors. The resulting dynamics may lead to the formation of a complete network with higher than Nash equilibrium effort levels. We summarize these points in Hypothesis 1.

**Hypothesis 1.** *In the Link benefit treatment, players compete for status by increasing their own effort levels in order to attract more incoming links. Accordingly, more links will be formed which* *further increases effort provision. The network converges to the complete network with overprovision of effort relative to the Nash equilibrium level.*

In the second status setting, we provide information on each individual's relative ranking in terms of payoffs earned within the group. This information is provided on the feedback page at the end of each period. We refer to this treatment as *Ranking*. Again, theoretically, the ranking information should have no impact on the individual's effort and linking decisions. However, as shown in the literature (e.g., Charness et al. 2014), individuals compete for achieving a higher rank in the group even if the ranking itself does not have any payoff consequence. They may attribute additional utility to doing better than others in the group. However, how to achieve a higher payoff than others is unclear. On the one hand, players can increase their own effort levels and initiate more links to others with high effort levels to have a higher payoff. This is because the benefits from a link increase with the effort levels of the two individuals involved in the link.[4] This results in a positive effect on effort provision and network connectivity (Mas and Moretti, 2009), which may lead to the formation of an efficient complete network. On the other hand, players who value status sufficiently highly or have strong competitive preferences, may sabotage others by withdrawing links to them (Carpenter et al., 2010; Charness et al. 2014). Both players would receive lower payoffs in this way, but the person who saves the linking costs will reduce payoffs less and may achieve higher relative ranking.[5] In turn, having less links discourages effort provision, pulling the group towards the empty network. These conjectures lead us to the following alternative hypotheses.

**Hypothesis 2a.** *In the Ranking treatment, to achieve a higher-ranked position, players may increase their own effort levels and initiate more links. The network may converge to the complete network with higher than equilibrium effort level.*

---

[4] A link between individuals $i$ and $j$ is beneficial to add if $0.4x_i x_j > 3.9$. This follows from equation (1).
[5] If individual $i$ cuts a link to individual $j$, $i$ looses $0.4x_i x_j - 3.9$ while individual $j$ looses $0.4x_i x_j$.

**Hypothesis 2b.** *In the Ranking treatment, to achieve a higher-ranked position, players may sabotage others by initiating less links which induces lower effort provision. Status competition may lead to lower effort provision and less connections being formed.*

In our fourth treatment, we include both types of social status manipulations: individual receive benefits from incoming links and the payoff ranking information is also provided. This treatment is especially interesting if Hypotheses 1 and 2b will be confirmed in which case the two manipulations yield opposite results. In this case, the *Interaction* treatment allows us to study which effect dominates the other.

## 2.3 Experimental procedures

The experiment lasted for 40 rounds. At the beginning of the experiment, participants were randomly divided into groups of 5. Groups remained unchanged throughout the experiment. Since status competition is more salient when individuals repeatedly interact with each other, we opted for the partner matching to establish a fixed set of peers.

At the beginning of each round, players engaged in the network formation game as described above with the other 4 players within the same group. More precisely, every player simultaneously and independently chose an effort level $e_i \in [0, 20]$ and the players to whom she would like to link to. Afterwards, the network was generated and the points earned by each individual were calculated according to the payoff function in equation (1) or (2), depending on the treatment.

At the end of each round, players were provided with detailed feedback on the outcomes of the round. These include her effort level chosen, her network neighbors, the list of other group members she initiated a link to and received a link from, her payoffs in total and its breakdown (i.e., the benefits from effort choice, the cost of making effort, and the cost of initiating links); a graph of the network generated in the group, and a table with information on other group members: their effort choices and lists of network neighbors. For the *Link benefit* treatment, the breakdown also contained information about the benefit obtained by receiving links from others. For the *Ranking* treatment, the players were also informed about their own payoff rank in the group, and information on the ranking of other group members was listed in the information table. The *Interaction* treatment contains full information on all these outcomes. The experimental instructions attached in the Supplementary Materials provide screenshots of the decision and feedback screens.

The payoff of the network game for each player was the sum of per round payoffs over all the 40 rounds.[6] Before playing the network game, a player was asked to play 5 practice rounds which were not counted in the payoffs.[7] At the end of the network game, there was a questionnaire containing three components: 1) demographic information on gender, age, country of origin, year of study, major, and also feedback on the strategy used in the network game; 2) an incentivized cognitive-reflection test (Frederick, 2005); and 3) an incentivized risk-aversion measure (based on Gneezy and Potters, 1997).[8] Accordingly, the total payoff from the experiment contains the sum of payoffs in the network formation game and the two incentivized components in the questionnaire. These earnings in terms of points were converted to real money using the exchange rate 1 RMB = 15 points. In addition to this amount, each participant received a show-up fee of 20 RMBs.

The experiment was conducted in the Experimental Economics Laboratory at Xi'an Jiaotong-Liverpool University.[9] The subjects were recruited from a pool of undergraduate and graduate students via the Online Recruitment System for Economic Experiments (ORSEE) platform (Greiner, 2015). 60.5% of the participants were female and 96% of them are of Chinese nationality, the average age in the sample was 20.925. The experiment was computerized using oTree (Chen et al., 2016). After reading the instructions, the participants had to answer a series of control questions before going to the 5 practice rounds and then the 40 rounds of the incentivized network game. The experimental instructions are provided in the Supplementary Materials. Overall, we recruited 10 groups per treatment therefore 10*5*4=200 subjects participated in the study. A typical session lasted for 120-150 minutes on average, with mean total earnings of 92.37RMBs (including 20RMB of show-up fee). The experiment was registered in the AEA RCT Registry (Horvath and Jia, 2023).

---

[6] Note that in principle, the total payoffs over the 40 rounds can be negative since the per round payoff can be negative with the given payoff function in equation (1). In this case, the player would receive a payoff of zero from the network formation game. However, this never happened in practice.

[7] During the practice rounds, the participants played against the computer that chose random strategies that were fixed for the 5 rounds to facilitate the learning of trade-offs embedded in the decisions.

[8] On average, individuals answered 2.4 questions correctly in the cognitive reflection test consisting of 3 questions. In the investment task, individuals could invest 40 points into a lottery that paid 0 and 2.5 times the investment with equal probabilities. The average amount invested was 20.42.

[9] The experiment was administered in English, which is the language of instructions at the university where the experiment took place.

## 3. Results

### 3.1 Network structure

We start describing the experimental results by analyzing the network structures formed during the experiment. We compare the networks formed to the theoretical predictions and study the impact of social status on the connectedness of the network. Table 3 reports the relative frequencies of equilibrium networks formed in the four treatments, considering all rounds and the last 10 rounds of the experiment, separately. We can observe that the empty and star networks are almost never formed during the experiment, and the complete network occurs most frequently among the three equilibrium network structures. However, the complete network is formed only in less than 15% of the times. Therefore, we can conclude that in general, equilibrium networks are rarely formed in the experiment.[10]

Table 4 shows summary statistics of the network structures formed during the experiment, including the number of links, the average number of neighbors (degree) and the minimum and maximum degrees in the four treatments. Considering the last 10 periods of the Baseline treatment, on average 7.25 links were formed in the group out of the maximum possible 10 links. The average degree was 2.937, while there was considerable heterogeneity in connectedness among the group members. This is captured by the difference between minimum and maximum degrees in the group, which are 2.28 and 3.65 on average, respectively.[11] To evaluate our hypotheses about the impact of social status on the number of links formed, we compare the network statistics between the Baseline and the other three treatments by Mann-Whitney rank-sum tests. As the results in Table 4 indicate, we find no statistically significant treatment differences for any of the network measures. This holds regarding the network statistics both for all periods and for the last 10 periods only. Our hypotheses 1 and 2a stating that introducing status manipulations to the game would increase the connectedness in the group are thus falsified by the experimental results. This is summarized in our first result.

---

[10] Moreover, comparing the treatments using chi-square tests, we find that the relative frequencies of equilibrium networks are not statistically significantly different across the treatments. Results omitted for brevity.

**Result 1.** *Introducing status manipulations to the network formation and effort provision game has no significant impact on the network structure as captured by the number of links, average, minimum, and maximum degrees.*

**Table 3: Relative frequency of equilibrium networks**

|  | All periods | | | Last 10 periods | | |
| --- | --- | --- | --- | --- | --- | --- |
| **Treatment** | **Complete** | **Empty** | **Star** | **Complete** | **Empty** | **Star** |
| **Baseline** | 0.090 | 0 | 0.008 | 0.140 | 0 | 0.010 |
| **Link Benefit** | 0.128 | 0 | 0.003 | 0.130 | 0 | 0 |
| **Ranking** | 0.085 | 0 | 0.003 | 0.080 | 0 | 0.010 |
| **Interaction** | 0.100 | 0 | 0 | 0.100 | 0 | 0 |

*Note:* To compute the relative frequency, we count the number of times a given equilibrium network was formed and we divide it by the number of periods.

---

[11] In Figure A1 in the Appendix, we show the evolution of average degree over the 40 periods of the experiment. We also depict the degree distributions in the four treatments in Figure A4.

Table 4: Network statistics (averages)

| Treatment | All periods | | | | Last 10 periods | | | |
|---|---|---|---|---|---|---|---|---|
| | Number of links Mean (Std. dev.) | Average degree Mean (Std. dev.) | Maximum degree Mean (Std. dev.) | Minimum degree Mean (Std. dev.) | Number of links Mean (Std. dev.) | Average degree Mean (Std. dev.) | Maximum degree Mean (Std. dev.) | Minimum degree Mean (Std. dev.) |
| Baseline | 6.878 (1.598) | 2.798 (0.594) | 3.595 (0.400) | 2.050 (0.673) | 7.250 (1.928) | 2.937 (0.738) | 3.650 (0.486) | 2.280 (0.855) |
| Link Benefit | 7.683 (1.073) | 3.076 (0.429) | 3.808 (0.199) | 2.288 (0.635) | 7.770 (1.087) | 3.111 (0.432) | 3.830 (0.177) | 2.360 (0.654) |
| Ranking | 7.315 (1.112) | 2.944 (0.421) | 3.660 (0.287) | 2.193 (0.510) | 6.740 (1.087) | 2.735 (0.432) | 3.550 (0.177) | 1.980 (0.654) |
| Interaction | 7.715 (0.789) | 3.093 (0.306) | 3.840 (0.186) | 2.280 (0.453) | 7.410 (1.149) | 2.975 (0.452) | 3.780 (0.274) | 2.110 (0.697) |
| Treatment comparisons, MW-test, test statistic (p-value) | | | | | | | | |
| Link benefit vs. Baseline | -0.907 (0.364) | -0.907 (0.364) | -1.512 (0.131) | -0.605 (0.545) | -0.302 (0.762) | -0.302 (0.762) | -0.113 (0.909) | -0.113 (0.909) |
| Ranking vs. Baseline | -0.643 (0.521) | -0.529 (0.597) | 0.038 (0.969) | -0.378 (0.705) | 0.567 (0.571) | 0.680 (0.496) | 0.643 (0.521) | 0.756 (0.449) |
| Interaction vs. Baseline | -1.058 (0.289) | -0.983 (0.326) | -1.814 (0.069) | -0.756 (0.449) | -0.756 (0.939) | 0.151 (0.879) | -0.454 (0.650) | 0.302 (0.762) |

*Note:* We compute network measures for each group and each period, including the number of links, average degree, minimum and maximum degrees. We take averages of these over all or the last 10 periods for each group to form independent observations and compare them across treatments by Mann-Whitney non-parametric tests.

### 3.2 Effort provision and payoffs

Next, we consider effort provision and payoffs in the four treatments. We provide summary statistics of these variables from the last 10 periods of the experiment and statistical tests comparing treatments in Table 5. For the tests, we create independent observations by averaging over the last 10 periods and the 5 group members, which gives one observation per group. The average effort level in the Baseline treatment was 4.494. We compare this value to the average optimal effort on the networks formed in the experiment in the last 10 periods, which can be computed based on the Katz-Bonacich centrality.[12] We find that the average effort in the experiment (4.494) is significantly higher than the average optimal effort level of 3.591 (MW-test, z-score: 2.644, p-value: 0.008). Individuals thus overprovide effort relative to the optimal level which concurs with previous experimental evidence on this game (Gallo and Yan, 2021; Horvath, 2023). The average per period payoffs at 21.049, however, are relatively low compared to what could be obtained in the complete network in the equilibrium, which is 26.92 (see Table 1). The difference is statistically significant (one-sample Wilcoxon signed-rank test, z-score: 4.0, p-value: 0.017). Individuals thus do not exploit all gains of collaboration because they do not form sufficiently many links which would create positive spillovers in effort.

---
[12] We compute the optimal effort levels for each period, group and individual, then take average of those across periods and individuals, which gives one independent observation per group.

We compare effort provision and payoffs in the treatments with social status to the Baseline by non-parametric tests. We find that in the Link benefit treatment, the average effort level is 5.558, which is statistically significantly higher than the average effort level of 4.494 in the Baseline (MW-test, z-score: -2.495, p-value: 0.012). We also obtain that the average per period payoffs are higher in the Link benefit treatment than in the Baseline: 33.469 vs. 21.049, the difference is statistically significant (MW-test, z-score: -3.401, p-value: <0.001). This finding can be explained by two reasons: 1) the Link benefit treatment has higher average effort level which leads to higher efficiency; 2) in the Link benefit treatment, individuals receive extra payoffs for incoming links from the experimenter. To understand the source of higher payoffs, we adjust the payoffs in the Link benefit treatment by deducting the extra points received for incoming links for each individual and recomputing the average payoffs. We find that the adjusted average payoffs are 21.385 which are not significantly different from the payoffs in the Baseline treatment. The higher payoffs in the Link benefit treatment can thus be explained by the extra payoffs received for incoming links and higher effort provision does not lead to larger payoffs on average, perhaps because it has no impact on the number of links formed (see Result 1).[13] These results partially confirm our Hypothesis 1.

Turning to the comparison between the Ranking and Baseline treatments, we find that the average effort level in the Ranking treatment is 3.601, which is statistically significantly lower than the average effort of 4.494 in the Baseline treatment (MW-test, z-score: 2.041, p-value: 0.041). Providing information on payoff ranking thus leads to lower effort provision. It follows that the payoffs are lower in the Ranking than in the Baseline treatment, the average per period payoffs are 21.049 in the Baseline and 17.756 in the Ranking treatment. The difference is, however, not statistically significant at conventional significance levels (MW-test, z-score: 1.285, p-value: 0.199). Our results partially confirm Hypothesis 2b.

---

[13] Note that most missing links would be beneficial to add: they would create additional payoffs for the individuals involved and lead to positive spillovers from effort provision.

The two types of status thus have different effects on effort provision: while the benefits for incoming links increase effort, information on payoff ranking in the group reduces it. It is an interesting question which effect dominates when the two types of status operate at the same time in the Interaction treatment. We obtain that the effect of payoff ranking is dominant since the average effort level in the Interaction treatment is statistically significantly lower than the effort level in the Baseline (MW-test, z-score: 2.117, p-value: 0.034), while it is statistically indistinguishable from the average effort level in the Ranking treatment (MW-test, z-score: -1.058, p-value: 0.289). We also find that, after adjusting the per period payoffs for the benefits received for incoming links in the Interaction treatment, the payoffs are not statistically significantly different between the Interaction and Ranking treatments or the Interaction and Baseline treatments. We thus find that the lower effort levels induced by the payoff ranking information in the Interaction treatment do not lead to lower payoffs.

We further verify the results obtained at the group level by studying the treatment differences at the individual level. We regress the different outcomes on treatment dummy variables using multi-level random-effects regressions with standard errors clustered at the group level. The omitted category is the Baseline treatment. We control for individual characteristics, including gender, age, study year, risk preferences and cognitive-reflection test results. The results are shown in Table A1 in the Appendix. The findings confirm that there are no significant treatment differences in the number of links initiated, the number of links realized and the adjusted per period payoffs. We obtain that effort levels are significantly higher in the Link Benefit treatment, and significantly lower in the Ranking and Interaction treatments compared to the Baseline.

The following result summarizes our main findings in this section.

**Result 2.** *Benefits received for incoming links lead to higher effort provision, while having information on payoff ranking in the group leads to lower effort provision compared to the Baseline treatment. The latter effect dominates the former when both linking benefits and ranking information are received in the Interaction treatment. We find no significant treatment difference in payoffs.*

Table 5: Effort and payoffs in the last 10 periods

| Treatment | Effort level Mean (Std. dev.) | Optimal effort level on the network formed Mean (Std. dev.) | MW-test comparing effort to optimal Test statistic (p-value) | Payoffs Mean (Std. dev.) | Adjusted payoffs after deducting linking benefits Mean (Std. dev.) |
|---|---|---|---|---|---|
| **Baseline** | 4.494 (0.728) | 3.591 (0.363) | 2.644*** (0.008) | 21.049 (5.706) | 21.049 (5.706) |
| **Link Benefit** | 5.558 (0.817) | 3.674 (0.222) | 3.780*** (<0.001) | 33.469 (6.355) | 21.385 (4.284) |
| **Ranking** | 3.601 (0.828) | 3.491 (0.355) | 0.756 (0.940) | 17.756 (5.594) | 17.756 (5.594) |
| **Interaction** | 3.808 (0.496) | 3.603 (0.229) | 0.605 (0.545) | 29.434 (4.840) | 17.29 (2.745) |
| | Treatment comparisons, MW-test, test statistic (p-value) | | | | |
| **Link benefit vs. Baseline** | -2.495** (0.012) | | | -3.401*** (<0.001) | -0.226 (0.821) |
| **Ranking vs. Baseline** | 2.041** (0.041) | | | 1.285 (0.199) | 1.285 (0.199) |
| **Interaction vs. Baseline** | 2.117** (0.034) | | | -2.948*** (0.003) | 1.285 (0.199) |
| **Interaction vs. Ranking** | -1.058 (0.289) | | | -3.250 (0.001) | -0.605 (0.545) |

Note: We compute average effort and per round payoffs for each group and each period. We take averages of these over the last 10 periods for each group to get independent observations and present the standard deviation of these group-level quantities in the parentheses. In the second column, we report the NE average effort in the network formed in the experiment using the formula of Katz-Bonacich centrality. We compare this to the average effort in the experiment by two-sample Mann-Whitney non-parametric tests. Stars attached to the numbers in the third column in the upper panel and the numbers in the lower panel indicate significance levels of these tests: *** 1%, ** 5%, *10%.

## 4. Mechanisms
### 4.1. The impact of link benefits

In this section, we aim to understand the mechanisms that lead to the observed treatment differences in effort provision. We start by analyzing the positive impact of benefits received for incoming links on effort provision. Our Hypothesis 1 was that individuals will increase effort provision in order to attract more incoming links and thus receive higher benefits. To evaluate this hypothesis, we analyze the factors that affect link formation, with particular attention to whether individuals are more likely to initiate links to high effort providers. We run individual random-effects logistic regressions with a binary dependent variable that captures whether an individual $i$ has initiated a link to another individual $j$ in the group in period $t$ (Yes is coded as 1 and No is coded as 0). Our regression is thus at the link level. The independent variables are the period number $t$ when the decision was submitted, the two effort levels in period $t-1$ of the individuals involved in the link, a dummy variable indicating whether the opponent, individual $j$ has initiated a link to individual $i$ in period $t-1$, the lagged dependent variable indicating whether individual $i$ has initiated a link to individual $j$ in period $t-1$, and the interaction term of these two dummy variables. The results, in the form of odds ratios, are shown in Table 6 separately for each treatment.

We find across all treatments that the likelihood of initiating a link is significantly and positively correlated to the opponent's effort level, confirming part of our hypothesis saying that higher effort levels attract more links. In addition, we find that link formation is guided by three types of behavioral factors: *inertia, reciprocity,* and *strategic behavior* to save linking costs. Inertia means that individuals keep initiating the same link across periods which is indicated by the larger than 1 and significant odds ratios of the lagged dependent variable in all regressions. Reciprocity means that participants are more likely to form a link if the same link was initiated in period $t-1$ by both parties involved in the link. This effect can be seen by the statistically significant and larger than 1 odds ratio of the interaction term between the lagged linking intentions of individual $i$ to $j$ and individual $j$ to $i$. This implies that reciprocated links (that is, those initiated by both parties) are more likely to be formed again in the next period. Regarding strategic behavior, we obtain in the Baseline and Ranking treatments that individuals exhibit strategic behavior in the sense that they are less likely to initiate a link in period $t$ to an opponent who unilaterally initiated a link to them in period $t-1$. This is shown by the significant and lower than 1 odds ratio of the dummy variable indicating whether a link was received from the opponent in period $t-1$. Individuals thus expect that others keep linking to them which allows them to save the linking costs by not initiating a link to these opponents. Interestingly, this sort of strategic behavior is not present in the Link benefit and Interaction treatments when individuals receive benefits from incoming links. The same independent variable is not significant in these treatments. This finding suggests that individuals in these treatments may be more likely to form reciprocated links, that are initiated by both parties involved, which allow them to share the benefits received for incoming links.

To provide evidence on this conjecture, we form a dummy variable that takes the value of 1 if the link is reciprocated, and the value of 0 if it was initiated only by one of the parties involved. We regress this variable on the treatment dummy variables and individual characteristics using individual random-effects logistic regressions. The regression results in Table 7 indicate that the odds ratios of the Link benefit and Interaction treatment dummies are above 1 and statistically significant. This means that in the treatments where individuals receive benefits for incomings links, it is more likely that a link is reciprocated compared to the Baseline treatment. In contrast, the odds ratio of the Ranking treatment dummy is much smaller and it is significant only at the 10% level.

These results explain why we do not observe more links being formed in the Link benefit treatment relative to the Baseline despite finding that effort provision is higher and that higher effort attracts more links. Although individuals initiate more links in accordance with our Hypothesis 1, these links create more reciprocated links that are sponsored by both parties involved. The extra link formation activity thus does not lead to more links being created but it results in stronger links that allow individuals to share the extra benefits received for incoming links. These stronger links may also induce higher effort provision as they enhance cooperation between connected individuals.

These results are summarized as follows.

**Result 3.** *In the Link benefit treatment individuals provide higher effort presumably to attract more links. While individuals do not establish more new network neighbors in this treatment, they* *create more reciprocated links which allow them to share the benefits of incoming links. In turn, these stronger links support higher cooperation in effort provision.*

Table 6: Determinants of link formation decision

| Dependent variable: Initiate a link in t (1=Yes, 0=No) | (1) Baseline Odds ratio | (2) Linking benefit Odds ratio | (3) Ranking Odds ratio | (4) Interaction Odds ratio |
|---|---|---|---|---|
| Initiated a link in t-1 | 3.256*** | 1.940*** | 2.233*** | 1.847*** |
|  | (0.472) | (0.290) | (0.403) | (0.323) |
| Received a link in t-1 | 0.440*** | 0.913 | 0.591*** | 0.840 |
|  | (0.076) | (0.194) | (0.085) | (0.146) |
| Initiated*received a link in t-1 | 1.483** | 1.361* | 1.456** | 1.785*** |
|  | (0.235) | (0.223) | (0.221) | (0.313) |
| Own effort in t-1 | 0.949* | 0.929*** | 0.967 | 1.036 |
|  | (0.029) | (0.021) | (0.052) | (0.035) |
| Opponent's effort in t-1 | 1.507*** | 1.466*** | 1.359*** | 1.216*** |
|  | (0.096) | (0.072) | (0.081) | (0.069) |
| Period | 1.001 | 0.995 | 0.989** | 0.997 |
|  | (0.004) | (0.004) | (0.004) | (0.005) |
| Constant | 0.103*** | 0.159*** | 0.314*** | 0.359*** |
|  | (0.034) | (0.053) | (0.103) | (0.092) |
| Number of observations | 7,800 | 7,800 | 7,800 | 7,800 |

*Note:* Individual random-effects logistic regressions at the link level with standard errors clustered at the individual level (reported in parentheses). Odds ratios are reported. Dependent variable: initiating a link (1 if the individual initiated a link to a given opponent, 0 otherwise). Stars indicate significance levels: *** 1%, ** 5%, *10%.

### Table 7: Treatment difference in the reciprocation of links

| Dependent variable | (1) | (3) |
|---|---|---|
| Reciprocated link (1=Yes, 0=No) | Odds ratio | Odds ratio |
| Link benefit | 1.781*** | 1.723*** |
|  | (0.271) | (0.262) |
| Ranking | 1.319* | 1.295* |
|  | (0.203) | (0.192) |
| Interaction | 2.193*** | 2.116*** |
|  | (0.340) | (0.353) |
| Study year |  | 0.936 |
|  |  | (0.083) |
| Gender |  | 1.147 |
|  |  | (0.127) |
| Age |  | 1.028 |
|  |  | (0.060) |
| Cognitive reflection test result |  | 0.849*** |
|  |  | (0.051) |
| Risk pref. |  | 0.996 |
|  |  | (0.005) |
| Period | 0.989*** | 0.989*** |
|  | (0.002) | (0.002) |
| Constant | 0.328*** | 0.319 |
|  | (0.041) | (0.352) |
| Observations | 23,672 | 23,672 |

*Note:* Individual random-effects logistic regressions at the link level with standard errors clustered at the individual level (reported in parentheses). Odds ratios are reported. Dependent variable: reciprocated link (1 if the link was initiated by both individuals involved in the link, 0 if the link was initiated only by one individual involved in the link). Omitted category is the Baseline treatment. Stars indicate significance levels: *** 1%, ** 5%, *10%.

### 4.2. The impact of payoff ranking information

Turning to our second treatment variation on social status, we aim to understand why providing information on relative payoff ranking in the group reduces effort provision. To this end, we analyze how the behavior of an individual in a given period $t$ is affected by her achieved ranking in the group in the previous period $t-1$. We regress the change in effort, the change in number of links initiated, the change in ranking in the group and the change in payoffs between $t$ and $t-1$ on the ranking obtained in the group in $t-1$. We split the group members by their payoff ranking into three groups: ranked in the top 2 (1st and 2nd), ranked 3rd in the group and ranked in the bottom 2 (4th and 5th). We add two dummy variables to the regressions, one representing the top, and the other the bottom 2 positions, which renders being ranked 3rd as the reference category. We use individual random-effects regressions with standard errors clustered at the individual level. The results are shown in Table 8 for the Ranking and Interaction treatments, separately.

We can see in column 1 of Table 8 that being ranked $4^{th}$ or $5^{th}$ in the previous period in the group in terms of payoffs induces individuals to exert less effort in the next period, while those ranked $1^{st}$ and $2^{nd}$ do not change their effort choices significantly. Individuals who decrease their effort, increase their payoffs as shown in column 4 of Table 8 by the significant and positive coefficient of the dummy variable representing low ranking in the group. Reducing effort increases payoffs because individuals provide higher effort than the optimal value in the early periods of the game, a smaller effort thus falls closer to the payoff maximizing choice. Due to this dynamics, the average effort decreases over time in the group, this can be observed in Figure A2 in the Appendix, which depicts the evolution of effort over time.

While low-ranked individuals increase their payoffs by reducing effort, high-ranked individuals earn lower payoffs due to the negative externalities that is exerted on them by their peers who provide less effort. As a consequence of these changes in payoffs, the ranking of individuals within the group reverses: low-ranked individuals climb to better relative positions, while high-ranked individuals decline in relative standing. This is shown by the estimation results in column 3 of Table 8 where the dependent variable is the change in payoff ranking within the group. All results mentioned so far are valid both for the Ranking and the Interaction treatments.

Regarding the impact of this dynamics on linking decisions, we obtain that individuals do not change the number of links initiated depending on their ranking in the group in the Ranking treatment. This is indicated by the insignificant coefficients of the ranking dummy variables in column 2 of Table 8. In the Interaction treatment, in contrast, we obtain that those ranked low in the group in the previous period, will initiate significantly less links in the next period. Note that in this treatment initiating a link creates benefits for the recipient of the link. Low-ranked individuals thus possibly intend to reduce the benefits received by others to improve their own relative position by reducing the payoffs of others in the group. We summarize these results in the following statement.

**Results 4.** *When payoff ranking information is provided, individuals ranked low in the group reduce their effort to increase their own payoffs and ranking in the group, which leads to lower payoffs and ranking for others in the group due to negative spillovers. The resulting dynamics leads to lower effort provision compared to the Baseline treatment.*

Table 8: The impact of payoff ranking within the group on change in behavior

| | \(1\) | \(2\) | \(3\) | \(4\) |
|---|---|---|---|---|
| | **Ranking treatment** | | | |
| | Change in effort between t and t-1 | Change in number of links initiated between t and t-1 | Change in ranking in the group between t and t-1 | Change in payoffs between t and t-1 |
| Low ranking in the group in t-1 (4th and 5th) | -0.298*** | -0.060 | -1.223*** | 3.917*** |
| | (0.056) | (0.085) | (0.083) | (0.439) |
| High ranking in the group in t-1 (1st and 2nd) | -0.010 | 0.091 | 0.935*** | -2.593*** |
| | (0.050) | (0.069) | (0.104) | (0.374) |
| Constant | 0.090** | -0.023 | 0.115 | -0.480 |
| | (0.038) | (0.057) | (0.095) | (0.307) |
| Number of observations | 1,950 | 1,950 | 1,950 | 1,950 |
| | **Interaction treatment** | | | |
| | Change in effort between t and t-1 | Change in number of links initiated between t and t-1 | Change in ranking in the group between t and t-1 | Change in payoffs between t and t-1 |
| Low ranking in the group in t-1 (4th and 5th) | -0.289*** | -0.118** | -1.073*** | 6.949*** |
| | (0.098) | (0.054) | (0.094) | (1.112) |
| High ranking in the group in t-1 (1st and 2nd) | -0.019 | 0.063 | 1.458*** | -8.021*** |
| | (0.048) | (0.043) | (0.103) | (0.757) |
| Constant | 0.103** | 0.023 | -0.154* | 0.727 |
| | (0.049) | (0.033) | (0.087) | (0.563) |
| Number of observations | 1,950 | 1,950 | 1,950 | 1,950 |

*Note:* Individual random-effects regressions with standard errors clustered at the individual level (reported in parentheses). Dependent variable: change in effort (column 1), change in number of links initiated (column 2), change in ranking in the group (column 3) and change in payoffs (column 4) between period t and t-1. Sample is the Ranking treatment in the upper and the Interaction treatment in the lower panels. Stars indicate significance levels: *** 1%, ** 5%, *10%.

## Conclusions

In this paper, we study the impact of competition for social status on the efficiency of collaboration in endogenous social networks. In our setting, individuals choose a costly collaborative effort and their network neighbors in each of the 40 rounds of our laboratory experiment. Direct neighbors benefit from each other's effort and their effort choices are strategic complements. These features are represented by the linear-quadratic payoff function in the experiment. Positive externalities in this game make the equilibrium outcomes social inefficient. We thus ask if a behavioral mechanism, namely, the competition for social status will facilitate the efficient choices.

In our Baseline treatment without social status, individuals overprovide effort relative to the Nash equilibrium level on the network that they formed. Despite this, payoffs are lower than the equilibrium payoffs in the most beneficial equilibrium of complete network because individuals do not form sufficiently many links. This result further underlines our motivation to increase efficiency in the network formation game. We introduce social status competition in two distinct forms. In the Link Benefit treatment, individuals receive additional monetary benefits for incoming links, which represent rents obtained from popularity. In our Ranking treatment, we provide information on relative standing in terms of payoffs within the group. We combine these two types of social status in the Interaction treatment.

We find that link benefits increase effort provision and strengthen the links in the network but do not lead to more links being formed. Consequently, the payoffs are not significantly higher after introducing this type of status to the Baseline treatment. Turning to the Ranking treatment, we find significantly lower effort provision compared to the Baseline treatment as individuals can improve their relative standing in the group by choosing a lower effort. The number of links and efficiency are, however, not significantly different in these two treatments. In the Interaction treatment, the impact of ranking information dominates the effects of link benefits, as we find that in this treatment the effort provision is significantly lower than in the Baseline.

Overall, we do not find any positive effect of social status competition on the efficiency of collaboration when network ties are endogenous. This is in contrast to the results obtained for the provision of local public goods in endogenous networks in van Leeuwen et al. (2020) who obtain that benefits received for incoming links create a virtuous competition which leads to the efficient provision of public goods. Our results invite further investigation regarding how to induce higher efficiency, in particular, via promoting link creation, in the network formation game with positive externality and complementarity of effort.

# Appendix

**Figure A1: The evolution of average degree in the four treatments over the 40 periods**

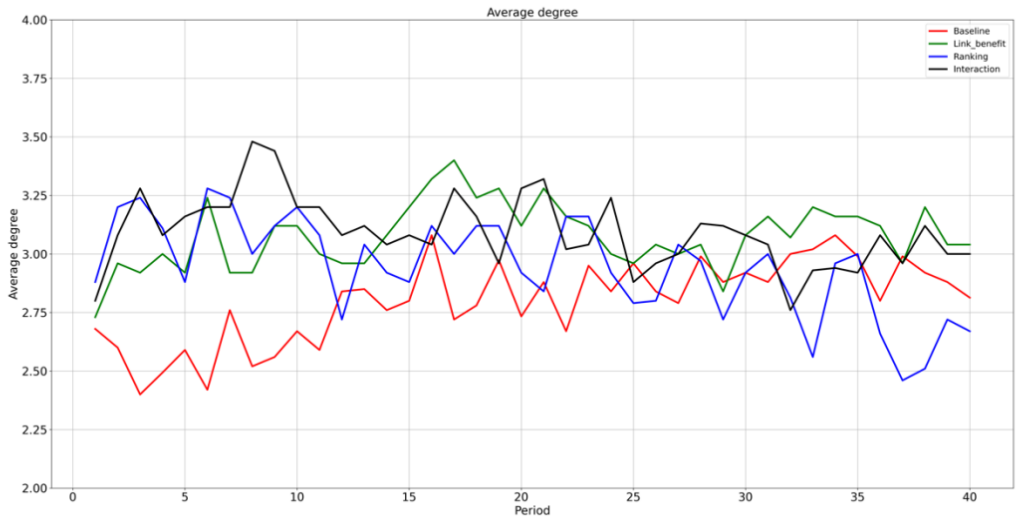

**Figure A2: The evolution of average effort level in the four treatments over the 40 periods**

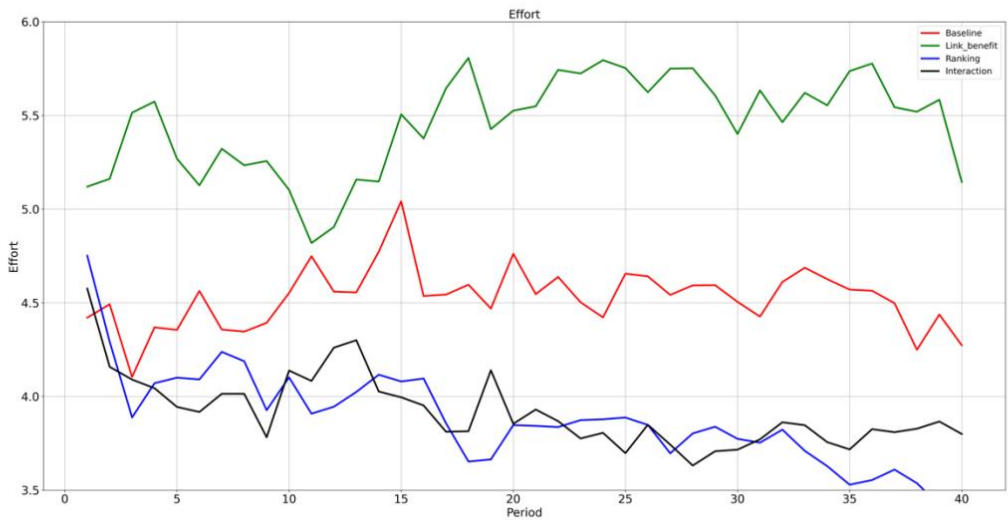

**Figure A3: The evolution of average payoffs in the four treatments over the 40 periods**

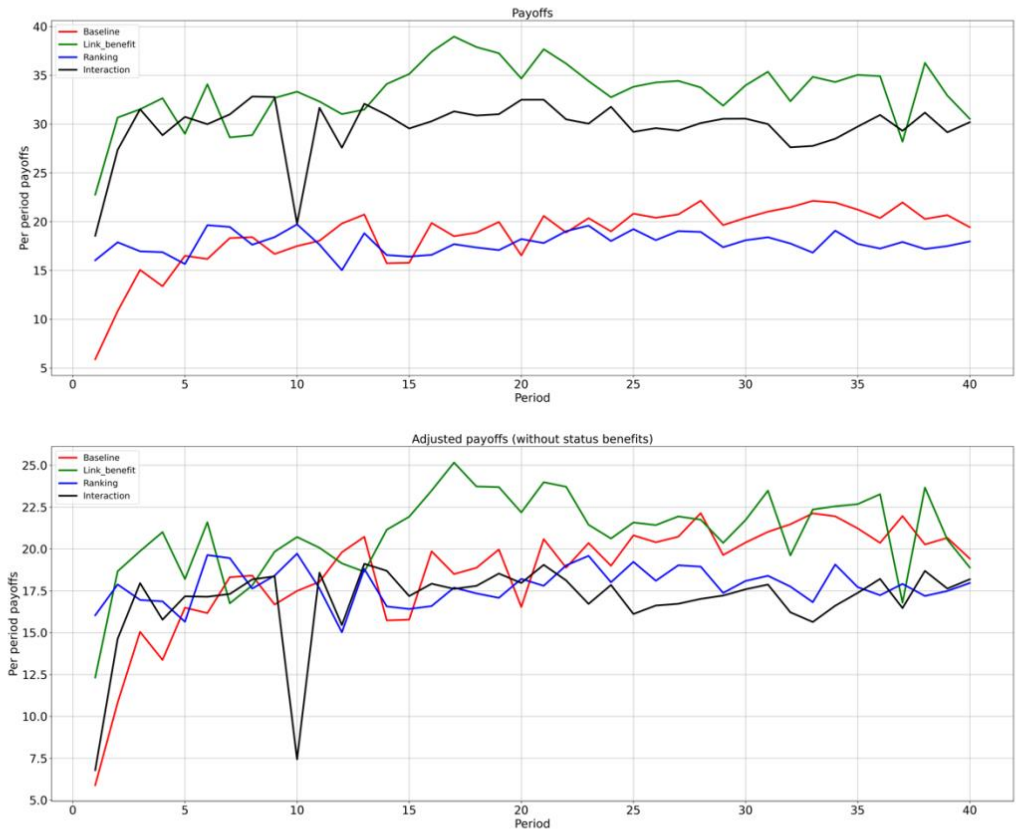

# Figure A4: Degree distribution in the four treatments

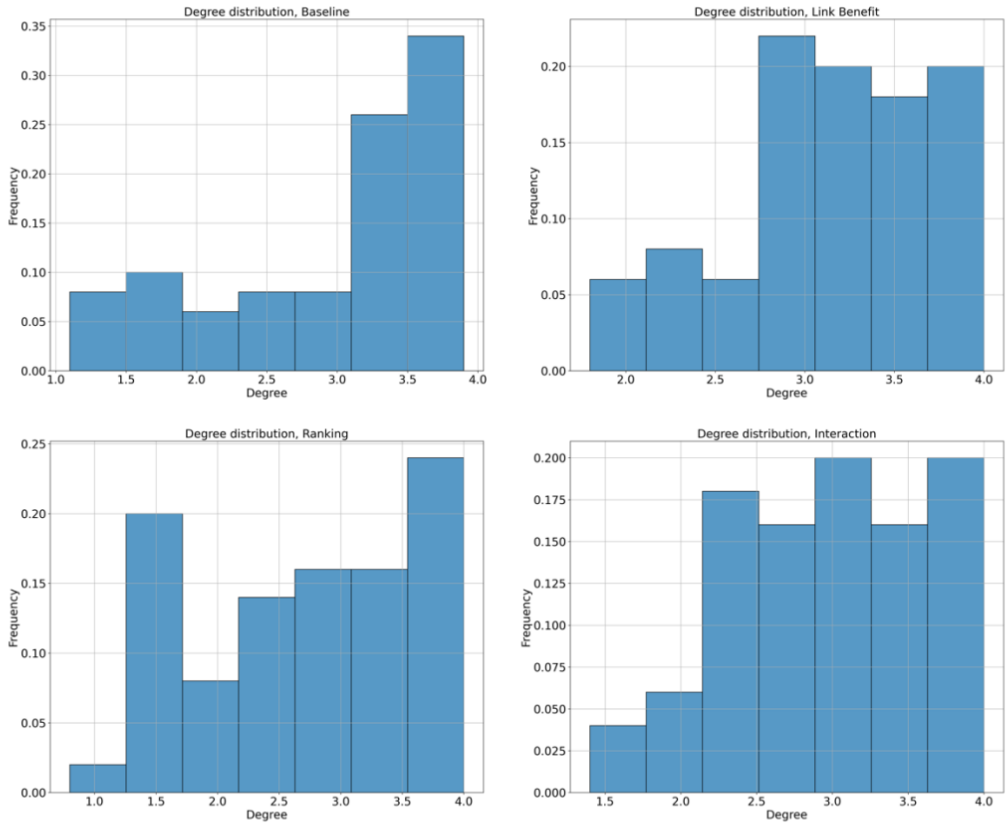

## Table A1: Treatment effects

|  | (1) Number of neighbors initiated | (2) Number of neighbors | (3) Effort | (4) Per period payoffs | (5) Adjusted per period payoffs |
|---|---|---|---|---|---|
| Link Benefit | 0.267 | 0.309 | 0.960*** | 14.905*** | 2.416 |
|  | (0.207) | (0.238) | (0.309) | (2.550) | (1.982) |
| Ranking | 0.140 | 0.170 | -0.644** | -0.755 | -0.782 |
|  | (0.179) | (0.237) | (0.302) | (1.886) | (1.872) |
| Interaction | 0.269 | 0.357 | -0.519* | 11.339*** | -1.891 |
|  | (0.206) | (0.223) | (0.284) | (2.142) | (1.865) |
| Period number | -0.004* | -0.001 | -0.004 | 0.087*** | 0.091*** |
|  | (0.002) | (0.003) | (0.004) | (0.028) | (0.026) |
| Gender (0=Male, 1=Female) | 0.175 | -0.007 | -0.046 | -1.173 | -0.962 |
|  | (0.121) | (0.060) | (0.090) | (0.982) | (1.015) |
| Age | 0.125** | 0.031 | -0.073 | -0.507* | -0.201 |
|  | (0.061) | (0.030) | (0.046) | (0.305) | (0.314) |
| Cognitive reflection test result | -0.167*** | -0.029 | 0.064 | 0.529 | 0.357 |
|  | (0.059) | (0.038) | (0.066) | (0.455) | (0.351) |
| Risk pref. | 0.002 | -0.001 | 0.001 | -0.004 | 0.006 |
|  | (0.005) | (0.003) | (0.005) | (0.035) | (0.034) |
| Study year | -0.233*** | -0.017 | 0.132* | 0.699 | 0.158 |
|  | (0.090) | (0.042) | (0.076) | (0.549) | (0.606) |
| Constant | -0.035 | 2.257 | 5.710*** | 25.463*** | 20.251*** |
|  | (1.204) | (0.655) | (0.915) | (6.761) | (6.886) |
| Number of observations | 8,000 | 8,000 | 8,000 | 8,000 | 8,000 |

*Note:* Multi-level mixed-effects regressions with standard errors clustered at the group level (reported in parentheses). Dependent variable: number of links initiated (column 1), number of neighbors (column 2), effort (column 3), per period payoffs (column 4), adjusted per period payoffs (column 5). Adjusted payoffs are formed by withdrawing payoffs received for incoming links in the Link benefit and Interaction treatments. The omitted category is the Baseline treatment. Stars indicate significance levels: *** 1%, ** 5%, *10%.